# Substrate-assisted nucleation of ultra-thin dielectric layers on graphene by atomic layer deposition

Bruno Dlubak, Piran R. Kidambi, Robert S. Weatherup, Stephan Hofmann, and John Robertson

*Department of Engineering, University of Cambridge, Cambridge CB2 1PZ, United Kingdom*

**We report on a large improvement in the wetting of $Al_2O_3$ thin films grown by unseeded atomic layer deposition on monolayer graphene, without creating point defects. This enhanced wetting is achieved by greatly increasing the nucleation density through the use of polar traps induced on the graphene surface by an underlying metallic substrate. The resulting $Al_2O_3$/graphene stack is then transferred to $SiO_2$ by standard methods.**

The growth of ultra-thin dielectric layers on graphene is central to unlocking several of its electronic applications. For example, RF transistors require thin high dielectric constant (high-K) gate oxide layers for good electrostatic control of the channel.[1–3] Lateral spin valves require electron transparent tunnel barriers to overcome the impedance mismatch and allow spin injection,[4,5] while bilayer pseudospin field effect transistors require a tunnel barrier to couple the two stacked graphene layers.[6] Interestingly, continuous 1nm sputtered $Al_2O_3$ films have already been grown on few-layer graphenes.[5] Technologically, it is desirable to use atomic layer deposition (ALD) to deposit dielectrics on graphene, but so far an additional seeding or interfacial layer is needed to yield continuous ALD oxide films 10nm thick, due to the poor wetting of graphene.[7–16] The seed layers can be either inorganic (1–5nm thick evaporated Al, Hf, Ti, or Ta layers oxidized in air before transfer to the ALD system[7,9–11]) or organic (spin-coated films of low-K polyhydroxystyrene derived polymer,[12] 2.5nm films of ozone activated poly-vinyl alcohol,[13] or perylene tetracarboxylic acid[14,15]). Indeed, the surface of a defect-free, clean graphene sheet is inert under the soft chemical ALD process.[7,8,14,16,17] Increasing the wetting of ALD dielectrics on graphene would thus allow a seed layer to be avoided. In this direction, Shin et al.[18] proposed to increase the surface energy of graphene by an $O_2$ plasma, which improves wettability, but at a cost of inducing defects in the graphene lattice.

In this paper, we demonstrate an alternative method to enhance wetting of ALD dielectric films on monolayer graphene without compromising its crystalline structure or seeding its surface. We show how to strongly improve wetting of graphene by tuning its surface energy using the substrate, as the graphene/metal proximity creates a dense network of polar traps and nucleation sites on the graphene surface. We focus experimentally on 10nm $Al_2O_3$ films grown by ALD on graphene materials supported on different substrates: graphite, graphene grown by chemical vapour deposition (CVD) transferred on $SiO_2$, CVD graphene grown on Cu,[19] and CVD graphene grown on a Ni-Au alloy.[20] The $Al_2O_3$ films are grown in a Cambridge Nanotech Savannah system using tri[methyl]aluminium as precursor and water as oxidant,[21] both carried in a 20 sccm flow of $N_2$ for 100 cycles, at growth chamber temperatures $T_{ch}$ leading to nominal 10nm thick $Al_2O_3$ films. In all presented experiments, precursors exposure doses are kept constant (1Torr, 0.015s pulses) and CVD-like reactions in the chamber are prevented by varying with temperature purges times between ALD half-reactions.[21]

For reference, we first grow 10nm $Al_2O_3$ films on highly oriented pyrolitic graphite (HOPG) at different $T_{ch}$ (80°C with 60s purges, 150°C with 20s purges, 200°C with 8s purges). The resulting films are characterized using a Hitachi S5500 scanning electron microscope (SEM). We found poor wetting of the $Al_2O_3$ on HOPG for all $T_{ch}$, as reported by others.[7,8,14,16,17] Films grown at lower temperatures have higher coverage, but are still clearly discontinuous. SEM images of the best-wetted films on HOPG (at $T_{ch}$=80°C) in Fig. 1(a) are clearly discontinuous, indicating an island-like Volmer-Weber growth mode.

We then investigate oxide ALD on CVD graphene. We focus in particular on high quality graphene (Raman D/G ratio ~0.05 at 532nm) grown on Cu at ~1000°C in controlled $Ar:H_2:CH_4$ atmospheres. The graphene sheet is placed on a $SiO_2$ substrate using a standard transfer process with a PMMA carrier film and a $FeCl_3$ Cu-etch and is then annealed in ultra-high vacuum (<$10^{-6}$ mbar). Compared to HOPG, wetting is expected to be better on transferred CVD graphene, as the transfer step can induce some defects. However, 10nm ALD $Al_2O_3$ films grown at 80–200°C on transferred CVD graphene also leads to poorly wetted films similar to those formed on HOPG (Fig.1(b)). Direct ALD of continuous ≤10nm $Al_2O_3$ films on HOPG or on transferred CVD graphene on $SiO_2$ is compromised.

We instead tried performing the ALD step directly onto the CVD graphene on the Cu foil step (Fig.2(a)), *before* the transfer. Importantly, the study is carried out on a similar piece of graphene/Cu foils as used previously for transferred graphene on $SiO_2$. Hence, the structural quality should be better on the un-transferred sheet and thus one might expect the wetting to be worse.[18,22] The ALD process is then repeated at each temperature. Surprisingly, the resulting $Al_2O_3$ films show dramatically improved wetting in the SEM. At $T_{ch}$=200°C, the coverage is already much improved compared to that on graphene/$SiO_2$, with coverage increasing from ~60% to >90%. The improvement is even more notable at $T_{ch}$=80°C (Fig.2(b)), as SEM reveals no structural defects in the $Al_2O_3$ film, in clear contrast to that found on graphite (Fig.1(a)) or on transferred graphene/$SiO_2$(Fig.1(b)). Concerning the growth of thinner oxide films by ALD, Fig.2(c) shows a 3nm $Al_2O_3$ film after 30 ALD cycles deposited at $T_{ch}$=80°C. Clearly, wetting of a 3nm $Al_2O_3$ film on graphene/Cu is greatly improved compared to wetting of a thicker 10nm film on graphene/$SiO_2$.

Many applications require the graphene film to be ultimately on an insulating substrate. An $Al_2O_3$/graphene layer, equivalent to that in Fig.2(b), was transferred on a $SiO_2$ substrate using a standard transfer process (Fig.3(a)). Significantly, the resulting $Al_2O_3$/graphene/$SiO_2$ stack can then be obtained over large areas following this scalable transfer process.[23] The structural quality of the graphene in this stack was checked using micro-Raman mapping (Renishaw InVia system) as shown in Fig.3(b). The observed uniform low intensity of the Raman D peak relative to the G peak in Fig.3(c) confirms that the $sp^2$ structure of the graphene is preserved,[24] even after ALD and transfer steps. The simple inversion of the order of the transfer and of the ALD growth steps allows us to benefit from the increased coverage of the $Al_2O_3$ film observed in Fig.2(b), providing a high quality encapsulated graphene layer.

To confirm the significance of our observations, complementary experiments are conducted at different $T_{ch}$ and on different graphene-like systems. In particular, monolayer graphene is also grown by CVD on a Ni-Au alloy, specifically tuned to allow growth of high quality monolayer graphene at a low (450°C) temperature, under a flow of $C_2H_2$.[20] Also, on a different set of samples, varying the exposure conditions yields the growth of multilayer graphene sheets (~5–10 layers) on either Cu or Ni-Au. Again, 10nm of $Al_2O_3$ are deposited by ALD on these samples at three different $T_{ch}$. All the observations made in terms of $Al_2O_3$ coverage for each $T_{ch}$ are reported in Fig.4(a). Two distinct trends are seen. A first set of experiments (on HOPG, monolayer/$SiO_2$, multilayer/Cu, and multilayer/Ni-Au) gives results corresponding to those in Fig.1: A decrease of $T_{ch}$ leads to improved coverage but with a poor wetting in all cases, the $Al_2O_3$ coverage remaining below 90%. A second set (on monolayer/Cu, monolayer/Ni-Au) gives results corresponding to those in Fig.2. The $Al_2O_3$ coverage is much improved compared to set 1 (above 90% in all cases) and films appear continuous for $T_{ch}$=80°C. We have observed similar trends for other ALD grown oxides (in particular, $HfO_2$ grown from water and tetrakis[dimethylamido]hafnium precursors).

We now consider the origin of the difference between the two experiments in Fig.4(a). We first estimate from Fig.1 the nucleation density ($N_n$) on HOPG or graphene/$SiO_2$ to be 2000–3000 per $\mu m^2$, while the Raman D/G ratio in Figs.3(b) and 3(c) corresponds to a defect density ($N_d$) of ~100–200 per $\mu m^2$.[25] As $N_d \ll N_n$, this indicates that defects may contribute to wetting,[18,22] but are not dominant. By contrast, when the surface is separated from a metallic substrate by only an atomically thin graphene layer, our coverage results suggest a nucleation density of ~$10^5$–$10^6$ per $\mu m^2$, and that other processes help increase wetting. The graphene/metal proximity has modified graphene's overall wettability.

One origin is the wetting "transparency" of graphene monolayers, noted by Rafiee et al.[26] These authors compared water contact angles on Si, Au, and Cu substrates with and without graphene to show that water adhered better on graphene when it is in intimate contact with a metallic substrate. This wetting transparency was also noted by Roos et al.[27] for bis(terpyridine) molecules adsorbed on graphene/Ru(0001). Moreover, Roos et al.[27] emphasised finite molecular adsorbtion on graphene/metal systems due to corrugations and direct interactions with graphene. In this direction, Pollard et al.[28] and Zhang et al.[29] showed by *ab initio* calculations and scanning tunnelling spectra that interaction between a monolayer graphene and an underlying metallic Rh(111) or Ru(0001) substrate creates a network of polar sites on the top surface, with lower adsoption energies for molecules like phtalocyanine, pentacene, or perylene tetracarboxylic diimide. This effect was previously seen for naphtahlocyanine and cuprate-phtalocyanine on a h-BN monolayer on Rh(111).[30–32]

Polar sites are a peculiarity of monolayer two-dimensional crystal/metal interactions.[27–32] They result from combined charge transfer (constrained by the difference of graphene and metal work functions) and wrinkles of the graphene sheet (constrained by the mismatch of the graphene crystal with the underlying metal). These studies suggest that the surface density of such polar traps ($N_t$) is ~$10^5$–$10^6$ per μm$^2$, i.e., $N_t > N_n$ for $N_n$ of HOPG or monolayer graphene/SiO$_2$.

The graphene transparency and the polar site network artificially modify the surface energy of the graphene sheet. This in turn alters the energy balance during the first ALD cycles from a water cluster formation/high desorption case[22] to a more homogeneous adsorption of water (see Fig.4(b)). This explains the transition from an island-like Volmer-Weber ALD growth (Fig.1) to a more two dimensional Frank-van der Merwe ALD growth mode (Fig.2) as water molecules adsorb more homogeneously over the graphene surface. Critically, the metal/graphene combination provides enhanced wetting effect without creating structural defects in the graphene 18,22 (Figs.3(b) and 3(c)).

While our experiments show that the effect of the substrate can result in a drastically higher Al$_2$O$_3$ coverage (from Fig.1 to Fig.2), we believe that a further enhancement is possible, and could allow better wetting at $T_{ch} > 80°C$. Indeed, it has been shown that the energetic landscape of the graphene sheet can be tuned by the selection of the metallic substrate: For instance, the trapping energy of the surface polar sites and their distribution on the surface as a regular superstructure template depend on the graphene/metal couple,[27–33] which may mean further optimisations of the water/surface interaction is possible. One could also use substrate assistance to provide a more controlled molecular coating of the graphene surface (Ref. 29 reports anisotropic trap energies forcing specific orientations of polar molecules), e.g., when using self-assembled monolayers as seeds for the ALD growth. Also investigation of graphene interactions with certain insulators may prove to lead to similar effects.

In summary, we have shown a dramatic increase in the wetting ability of ALD Al$_2$O$_3$ films by the assistance of a metallic substrate (Cu, Ni-Au) beneath the graphene layer, compared to thicker graphite layers or inert SiO$_2$. This assistance is conveniently fulfilled by the metallic catalysts commonly used during the CVD growth of graphene. This allows the nucleation of ultrathin Al$_2$O$_3$ films on clean graphene sheets, without introducing additional defects or seeds. The resulting Al$_2$O$_3$/graphene stack can then be routinely transferred onto a substrate of choice (here SiO$_2$). We highlight the role of the graphene/substrate interaction, which modifies the energetic landscape of the graphene surface and allows a more efficient trapping of water molecules during our Al$_2$O$_3$ growth by ALD. We envision this phenomenon to be also of importance for other controlled growths and coatings on graphene.

**FIGURES CAPTIONS**

**Figure 1.** ALD growth of 10nm $Al_2O_3$ at $T_{ch}$=80°C on (a) HOPG and (b) CVD graphene after transfer on $SiO_2$. SEM pictures taken in the same conditions at 3kV.

**Figure 2.** ALD growth of $Al_2O_3$ at $T_{ch}$=80°C on CVD graphene on Cu. (a) Large scale high quality monolayer graphene sheets grown on a Cu foil. Sample size is compared to a copper based 1 pence coin. SEM pictures taken in the same conditions at 3kV after the growth of (b) 10nm and (c) 3nm of $Al_2O_3$. The red arrow points to a selected <10nm wide crack.

**Figure 3.** (a) Process leading to well wetted $Al_2O_3$/graphene films on $SiO_2$. (b) Mapping of the Raman D/G ratio ($\lambda$=532nm, ×100 optic, 0.6μm spot diameter, and 1μm steps) of the encapsulated graphene sheet in-between $Al_2O_3$ and $SiO_2$ (c) Statistical analyses confirm that the D/G ratio is typically 0.05-0.06, similar to as grown graphene.

**Figure 4.** (a) Comparison of the coverage of the $Al_2O_3$ grown by ALD on different graphene-like materials and at different $T_{ch}$ (as observed through the SEM at 100k magnification). Two trends clearly emerge. Low coverage: HOPG, monolayer graphene on $SiO_2$ and graphene multilayers on Cu or Ni-Au; High coverage: monolayer graphene on Cu or Ni-Au. (b) Sketch of the trapping mechanism arising in the case of monolayer graphene/metal samples: water molecules are more efficiently adsorbed on the surface during the first ALD cycles and thus allow a more two dimensional Frank-van der Merwe growth of the $Al_2O_3$.

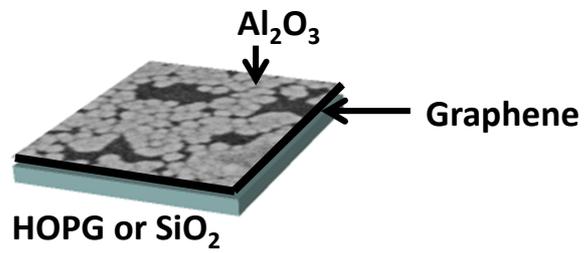

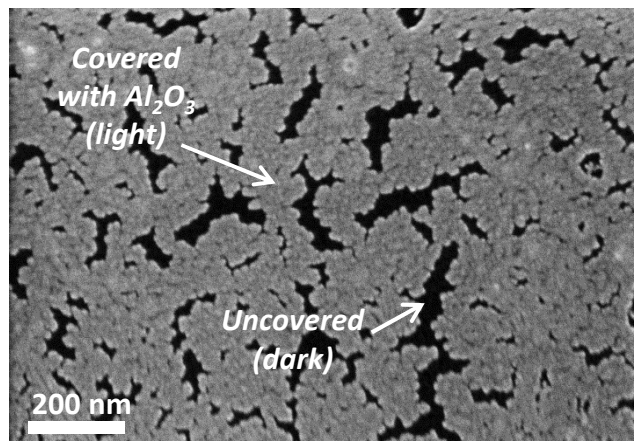

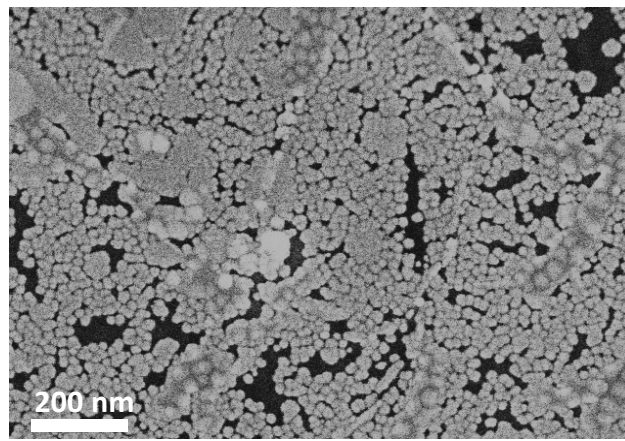

**Figure 1**

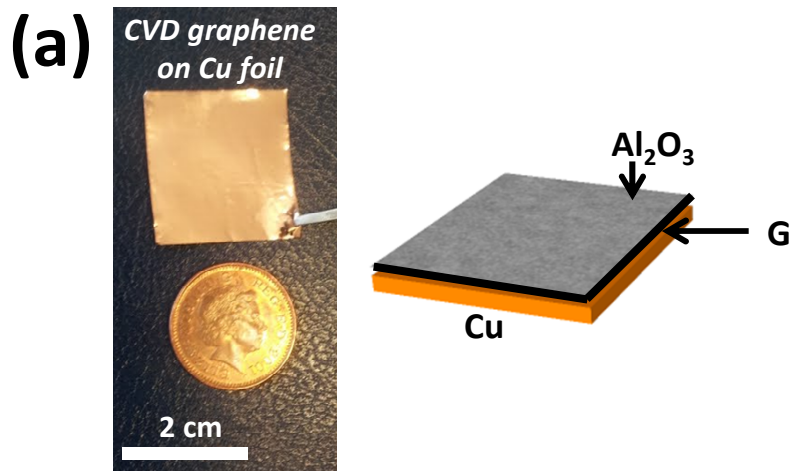

(a) CVD graphene on Cu foil

2 cm

Al$_2$O$_3$

G

Cu

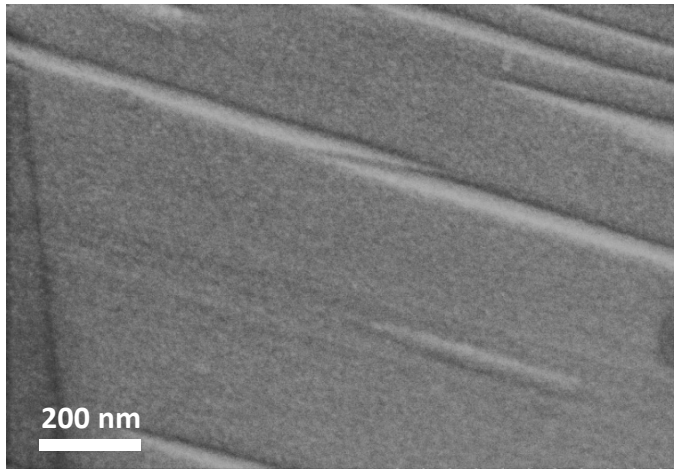

(b) 10 nm Al$_2$O$_3$/monolayer graphene/Cu

200 nm

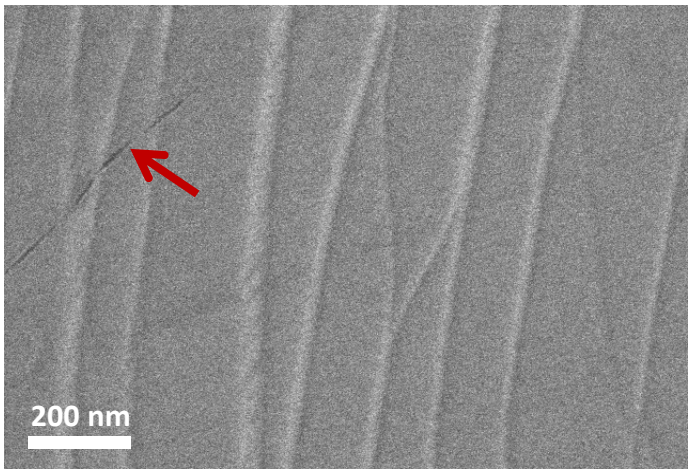

(c) 3 nm Al$_2$O$_3$/monolayer graphene/Cu

200 nm

**Figure 2**

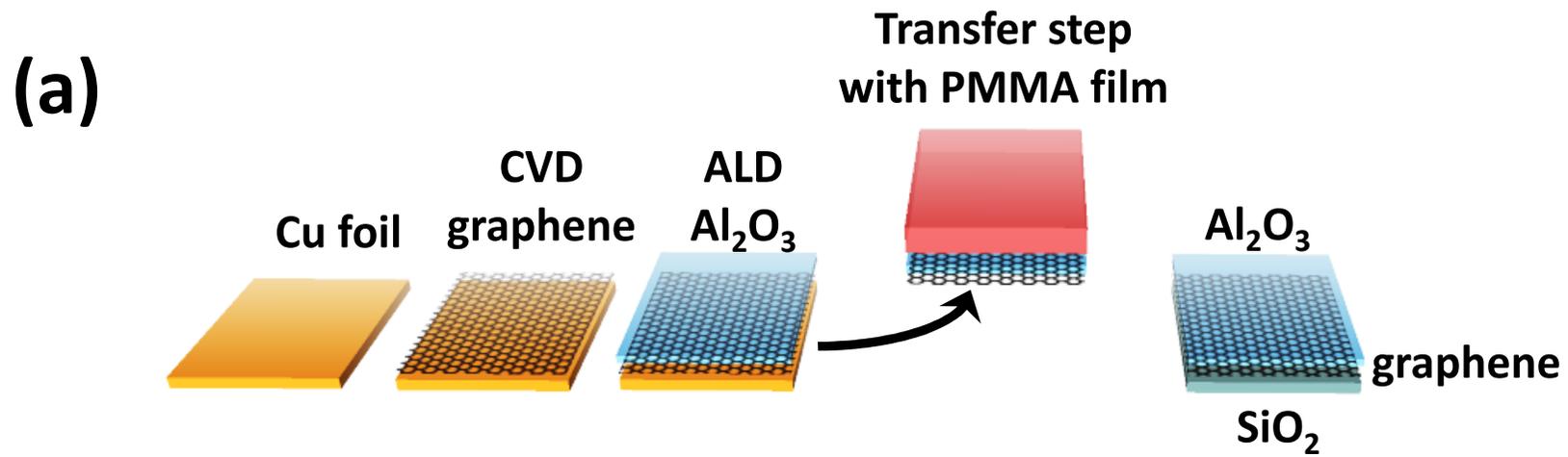
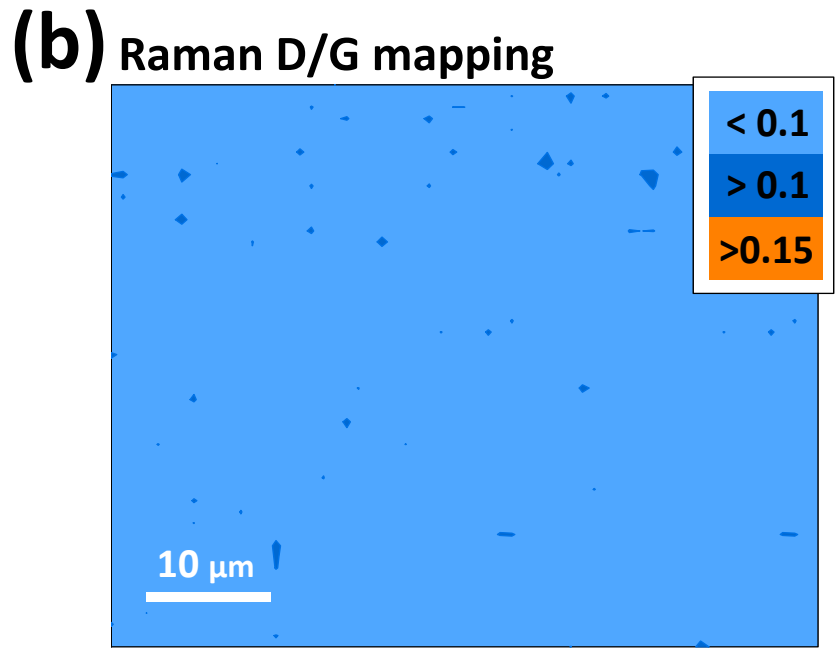
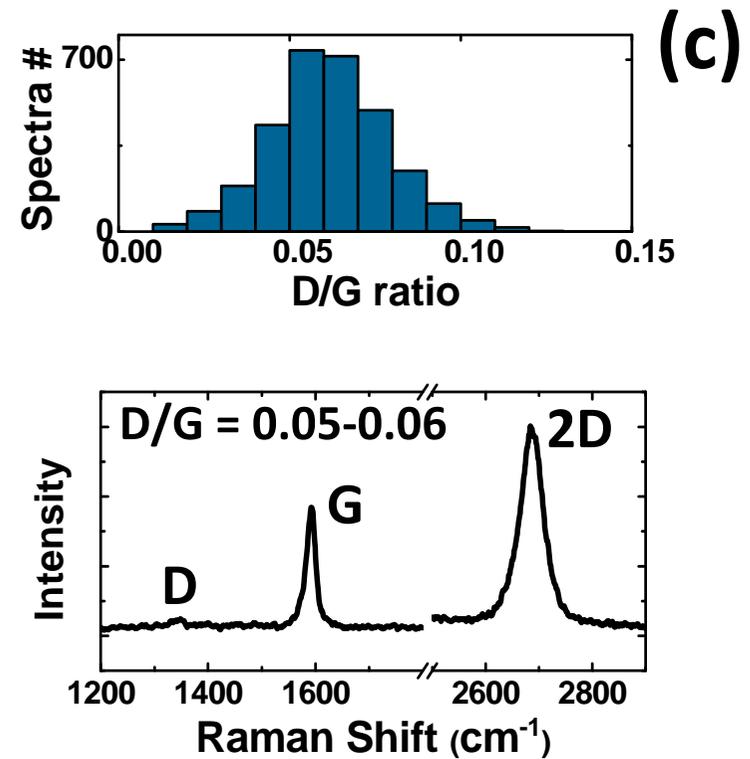

**Figure 3**

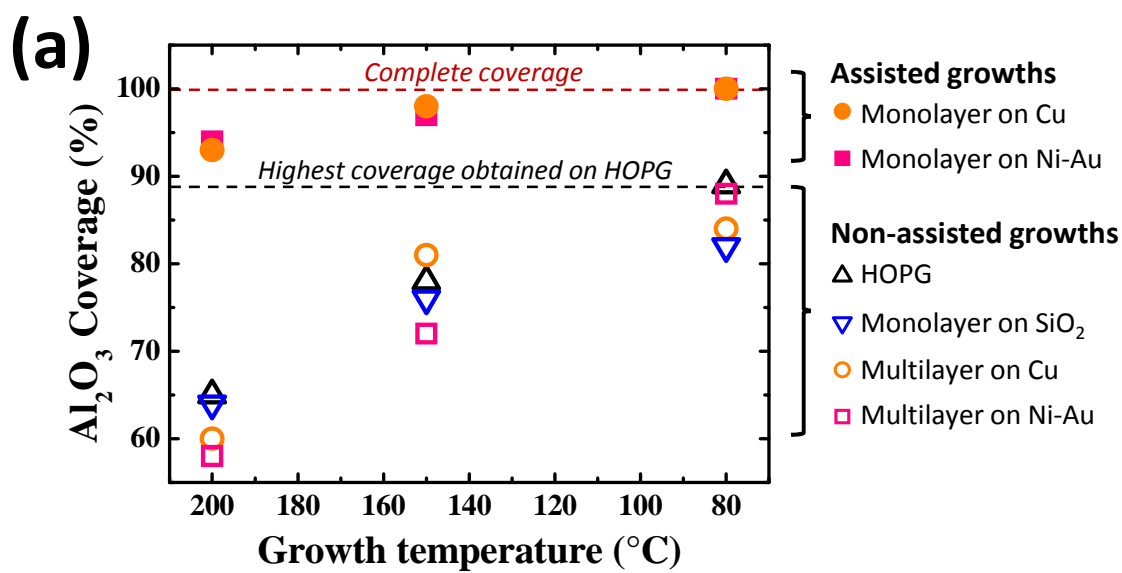

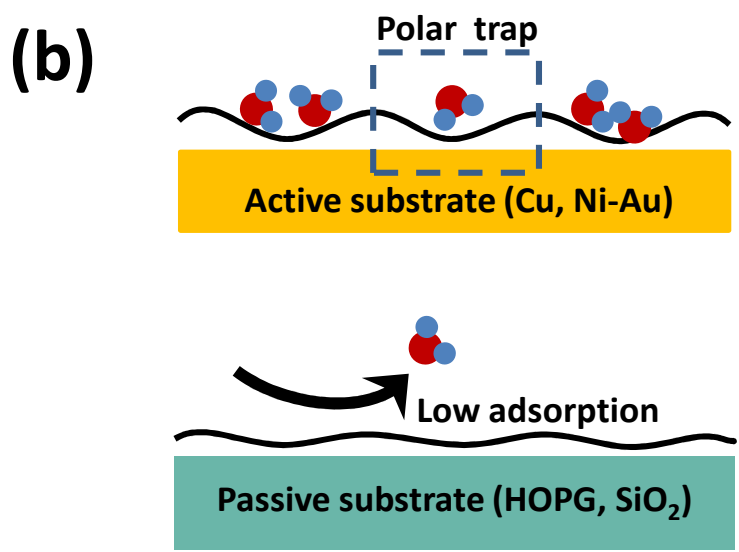

Figure 4